\documentclass[dvips]{article}
\usepackage{cite}
\usepackage{icrctc07}

\newcommand{\diff}[2]{\frac{d #1}{d #2}}
\newcommand{\pdiff}[2]{\frac{\partial #1}{\partial #2}}

\title{Particle acceleration by multiple parallel shocks}
\shorttitle{Particle acceleration by multiple parallel shocks}

\authors{Joni Tammi \& Paul Dempsey}
\shortauthors{Tammi \& Dempsey}
\afiliations{UCD School of Mathematical Sciences, University College Dublin, 
  Belfield, Dublin 4, Ireland.}
\email{joni.tammi@iki.fi, paul.dempsey@ucd.ie}

\abstract{We present both numerical and semi-analytical results on
  test-particle acceleration in multiple parallel shocks. We apply a
  kinetic Monte Carlo code and an eigenfunction expansion method to
  calculate the distribution functions for electron populations
  accelerated in subsequent parallel shocks with speeds ranging from
  non- to fully-relativistic. We examine the levels of particle
  anisotropy at the shocks and discuss the implications for AGN and
  microquasar jets.  }

\begin{document}
\maketitle

\section{Introduction}

Fermi acceleration at multiple non-relativistic shocks has been dealt
with in detail by \cite{MP93} and \cite{PM94}. However, many objects
that are likely to contain multiple shocks are relativistic, e.g., the
jets of active galactic nuclei (AGNi),\cite{Rees78} the internal
shock models of microquasars \cite{KSS00} and both the internal shock
and reverse shock models of gamma ray bursts.\cite{FWZ04,MR99} While
the external shocks in these sources are inevitably strong, internal
shocks can be weak, \cite{FWZ04,KSS00} either due to hydrodynamic
pressure in parallel shocks or magnetohydrodynamic pressure in
perpendicular shocks. Strong relativistic shocks are known to be
capable of producing hard particle spectra with $N(E)\propto E^{-2.2}$
via the first-order Fermi mechanism. On the other hand the first-order
mechanism produces much softer spectra with weak shocks.\cite{KGGA00}
However, in the case of a pre-existing hard power law population of
particles then either inclusion of the shock drift mechanism
\cite{BK90} or small-angle diffusion \cite{DD07} leads to an
amplification of this power law far above that expected from purely
adiabatic compression.

In this paper we consider hard power law spectra, \[ N(E)\propto
E^{-\sigma}, \] created at a strong shock by the first-order
acceleration mechanism, which are swept up by a weak shock. There is a
number of physical scenarios in which this is possible; one such
possibility is in microquasars and the parsec-scale jets of AGNi where
electrons accelerated at an external shock are advected downstream of
it towards the following internal shocks. Another scenario is the
collision of two expanding shells of plasma where the weak reverse
shock of the first shell sweeps up the high energy electrons produced
by strong forward shock of the second shell.

In this paper we restrict ourselves to mildly relativistic external
shocks, applicable to most AGNi and microquasars. However, the
internal shocks of these sources can still have high Lorentz factors
due to outbursts of the central engine.  

\section{Methods}

The semi-analytic eigenfunction approach is based on solving the
steady-state particle transport equation for the phase space
distribution function: 
\begin{eqnarray}
 \Gamma(u+\mu)\pdiff{f}{z}=\pdiff{}{\mu}D_{\mu\mu}\pdiff{f}{\mu},\nonumber
\end{eqnarray}
where $u$ is the bulk flow speed in the shock rest frame,
$\Gamma=(1-u^2)^{-1/2}$, $\mu$ is the pitch-angle and $D_{\mu\mu}$ is
the pitch-angle diffusion coefficient. This equation holds separately
upstream and downstream of the shock, and boundary conditions are
required to find the full solution. One condition is that the
distributions match at the shock front, i.e., 
\( 
 f_-(p_-,\mu_-,0)=f_+(p_+,\mu_+,0),
\) 
where the plus/minus sign denotes quantities upstream/downstream of the
shock. We also require the distribution to be bounded infinitely far
downstream. The distribution must be given far upstream and it is zero
when only injection at the shock is considered. We can expand $f$ as
\[
 f(p,\mu,z)=
 \sum_{i=-\infty}^\infty a_i(p) Q_i(\mu)\exp\left(\Lambda_iz/\Gamma\right)
\]
where $(Q_i(\mu),\Lambda_i)$ are eigenfunction--eigenvalue pairs satisfying
%
\[
 \diff{}{\mu}D_{\mu\mu}\diff{Q_i}{\mu}=\Lambda_i(u+\mu)Q_i.
\]
If the far upstream distribution is $g(p_-)$, then the boundary
conditions imply
\[
 g(p_-)+\sum_{i>0}a_i^{-}(p_-)Q_i^-(\mu_-)
 =\sum_{i\le0}a_i^+(p_+)Q_i^+(\mu_+). 
\] 
If the far upstream distribution follows a power law with index $s_1$,
i.e., $g(p_-)=A_0p_-^{-s_1}$, we can find solutions with
$a_i^{\pm}(p_\pm)=a_i^\pm p_\pm^{-s_1}$ by multiplying the matching
condition by $Q_j^+(u_++\mu_+)$ and integrating over $\mu_+$ for
$j>0$. The same applies also if we one considers only monoenergetic
injection at the shock. In this case, letting
\begin{eqnarray}
  W_{i,j}=
  \int_{-1}^{1}\left(1+u_{\rm rel}\mu_-\right)^{s_1}(u_++\mu_+)\times\nonumber\\
  Q_j^{+}(\mu_+)Q_j^-(\mu_-)\;d\mu_+, \nonumber
\end{eqnarray}
$s_1$ is found by the condition $\det{\bf W}=0$ as in \cite{KGGA00}.

\begin{table}
  \centering
  \begin{tabular}{|c|ccc|cc|}
    \hline
    Prof & $v_{\rm sh1}$ & $v_{\rm do1}$ & $r_{1}$ & $v_{\rm sh2}$ & $v_{\rm do2}$ \\ 
    \hline
    A & .3000 & .2293  & 3.95 & .7290  & .61933   \\
    B & .7070 & .5965  & 3.70 & .9770  & .95731   \\ 
    C & .9500 & .9105  & 3.25 & .9989  & .99783   \\ 
    D & .9800 & .9619  & 3.10 & .9997  & .99934   \\  
    \hline 
  \end{tabular}
  \caption[Flow Profiles Used]{\small The sets used.  The speed for
    the both shocks, $v_{\rm sh1}$ and $v_{\rm sh2}$, and the
    corresponding downstream flow speeds, $v_{\rm do1}$ and $v_{\rm
    do2}$, and the compression ratio of the first shocks, $r_1$, are
    given in the far upstream rest frame (observer's frame); all
    calculations and simulations are carried out in the relevant shock
    rest frames. Compression ratio of the second shock is 3 in all cases.}
  \label{table: profiles}
\end{table}

We also used a kinetic Monte Carlo test-particle
simulation\footnote{\texttt{
    http://www.iki.fi/joni.tammi/qshock}} for comparison. We injected
particles in the upstream of a one-dimensional step shock, and
followed them under the guiding-centre approximation until they reach
a pre-defined escape boundary. This boundary was set sufficiently far
away in the downstream to make sure the particles have reached
isotropy in the downstream plasma frame. When a particle crossed the
boundary, it was either ``absorbed to the downstream'', i.e., removed
from the simulation, or mirrored mimicking the case of the particle
recrossing the shock after returning from far downstream. This was
done with the help of the probability of return (see, e.g.,
\cite{JE91}). Also particle splitting was used to improve statistics:
when a particle reached certain energy, it was replaced by two
``daughter particles'', which were otherwise identical to their
``mother'', but had only half of the original statistical weight. 

In the simulation particles were scattered at the end of each Monte
Carlo time-step. The small-angle scatterings are elastic in the rest
frame of the scattering centres, which are taken to be moving with the
local flow speed, so the energy remains unchanged over a scattering in
the local flow frame.  Energy losses were omitted in these simulations
to allow for better comparison with the semi-analytical results.
\begin{figure}
  \centering
  \includegraphics[width=\columnwidth]{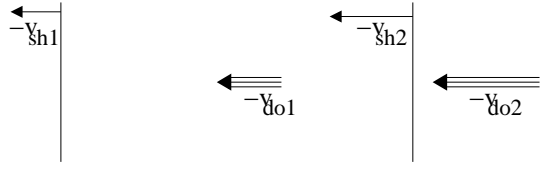}
  \caption[Two Shock Model]{\small Simple model of the flow profile
    considered in this paper. The shocks are far enough apart that
    particles once a particle crossed the second shock it can never
    return to the first shock.}
  \label{fig: multipleshock}
\end{figure}

For the first shock the injected particles have a small initial energy
and random direction. For the second shock we inject the particles
accelerated in the first shock. The particle properties were measured
in the local plasma frame, so the downstream properties for the first
shock simply become upstream properties for the second. However, the
the shocks in each case (see Table 1) were simulated separately, so
for example particles escaping upstream from the second shock cannot
travel back to the first shock.

\section{Results and Discussion}

As outlined in the introduction the process we are interested in the
process where a power-law distribution, pre-created at a strong shock,
is being swept and further accelerated by a weak shock.  We consider
four different velocity profiles (see Figure~\ref{fig: multipleshock}
and Table~\ref{table: profiles}), for which the compression ratio of
the first shock is calculated from the hydrodynamical jump conditions
for a plasma satisfying the J\"uttner-Synge equation of state. The
compression ratio of the second shock is 3 in all cases.
\begin{figure}
 \centering
 \includegraphics[width=1.0\columnwidth]{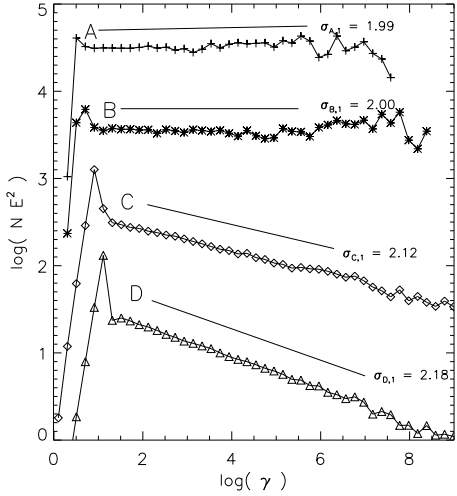}
 \caption[Energy Spectra Shock A]{\small Simulated energy spectra
   (multiplied by $E^2$) downstream of the first shocks for the
   profiles given in Table~\ref{table: profiles}. Spectra have been
   shifted vertically to allow for better comparison. Solid lines give
   slopes obtained from the eigenfunction method.}
 \label{fig: spectraA1d}
\end{figure}
\begin{figure}
  \centering
  \includegraphics[width=\columnwidth]{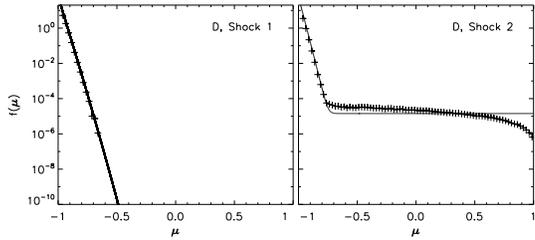}
  \caption[Comparison of $f(\mu)$ for the D case.]{\small Distribution
  of power-law tail particles at both D-profile shocks as measured
  in the upstream rest frame.}
  \label{fig: shockD1D2fmu}
\end{figure}

Energy spectra and pitch-angle distributions were obtained using both
the eigenfunction method and particle simulations, and there is good
agreement between the two methods. Figure \ref{fig: spectraA1d} shows
the energy spectra a few scattering lengths downstream of the first
shock for all profiles together with the semi-analytically obtained
spectral slopes. While the pitch-angle distribution is isotropic at
this distance downstream, it can be highly anisotropic at the shock
front.%
\footnote{This poses difficulties for comparing the $\mu$ distribution
  in the upstream, as for a relativistic shock (cf.~case D) one would
  need to simulate of the order of $10^{90}$ crossings to get one
  crossing with $\mu =+1$ due to the extreme anisotropy. For this
  reason, the data for simulated particles does not extend far from
  $\mu' = -1$ in this frame.} This can be seen in Figure \ref{fig:
  shockD1D2fmu}, showing the pitch-angle distribution at the both
  shocks for profile D, as measured in the corresponding upstream rest
  frame.
\begin{figure}
  \centering
  \includegraphics[width=0.8\columnwidth]{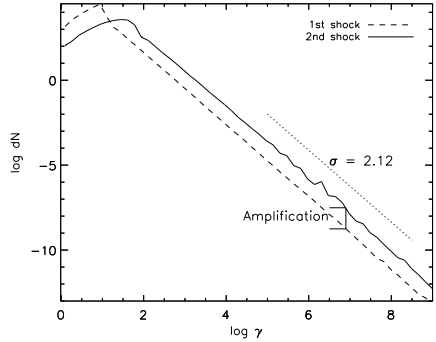}
  \caption{\small Particle spectra before (dashed line) and after the
    second shock (solid line), as measured in the local plasma frame,
    for the C case. The dotted line shows the slope obtained from the
    eigenfunction method.}
  \label{fig:shockD1D2}
\end{figure}

\begin{figure}
  \centering
  \includegraphics[width=0.8\columnwidth]{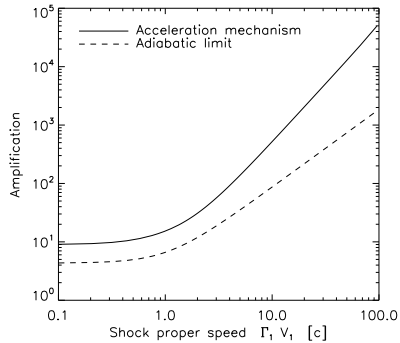}
  \caption[Comparison of $f(\mu)$ for the D case.]{\small
    Amplification of spectrum as a function of shock proper speed. See
    text for details.}
  \label{fig:gains}
\end{figure}

The spectral index doesn't change at the second shock crossing due to
the fact the the ``natural spectral index'' of such a weak shock is
greater than the index for the injected spectra. The power-law is
shifted due to the Lorentz transformation across the shock, and
additionally due to the further acceleration by the first-order
process. The compression of the plasma at the shock crossing doesn't
affect the plain energy spectrum, but needs to be taken into account
when comparing the distribution functions. In addition to 
compression and the change of frames across the shock, the shock
acceleration mechanism can lead to significant additional
amplification. Example is shown in Fig.~\ref{fig:shockD1D2}, where
the number of the power-law particles at a given energy (in the local
plasma frame) is increased by the factor of 17.4 (case C), and general
behaviour of the amplification with respect to the shock speed is
drawn for an example case (spectral index $\sigma=2$ and compression
ratio $r=3$) in Fig.~\ref{fig:gains}.

\begin{figure}
  \centering
  \includegraphics[width=0.8\columnwidth]{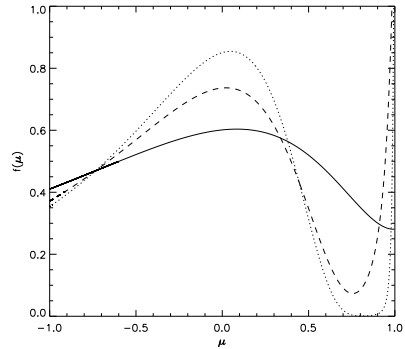}
  \caption[Comparison of $f(\mu)$ for the D case.]{\small
    Pitch-angle distribution, as measured in the downstream frame, of
    the particles for cases A (solid line), B, (dashed) and D (dotted)
    for the second shock.}
  \label{fig: fmu_ds}
\end{figure}

The peak near $\mu=+1$ in the particle pitch-angle distribution at the
shock measured in the downstream rest frame is due to particles that
cross the shock and never return. For the first shock these particles
are not considered as they are not in the power-law part of the
spectrum; in the second shock, however, these are well beyond the
thermal population and thus may have some signifigance e.g. in
radiation modelling, as the highest-energy particles can radiate their
energy away before they have had time to isotropise in the downstream
of the second shock. As expected, Figure \ref{fig: fmu_ds} shows that
this proportion increases as the shock becomes relativistic.

The eigenfunction method and test-particle simulations broadly agree
on the level of amplification. However, the eigenfunction method has
no cut-offs in particle energy due to assumptions about energy losses.
The simulations show that as well as an amplification in the magnitude
of the power law part of the distribution, the power law extends to a
higher cut-off energy. This will result in a higher peak energy in the
synchrotron spectra behind the shock than what exists ahead of the
second shock.

While this paper restricts itself to weak subsequent shocks, future
work will use both presented methods to examine re-accleration at
strong shocks in order to produce results comparible to \cite{MP93}
and \cite{PM94} in the relativistic limit.

In addition to energy losses due to, e.g., synchrotron emission or
adiabatic expansion, the current study omits possible effects due to
turbulence transmission (see Tammi, elsewhere in this volume),
second-order acceleration between the shocks \cite{VV05}, as well as
injection of new low-energy particles into the second shock.

\bibliographystyle{plain}

\begin{thebibliography}{3}\vspace{-0.9em} 
\bibitem{BK90} Begelman, M.~C., \& Kirk, J.~G.\ 1990, ApJ, 353, 66
\bibitem{BB96} Bicknell, G.~V., \& Begelman, M.~C.\ 1996, ApJ, 467, 597 
\bibitem{DD07} Dempsey, P., \& Duffy, P.~2007, Ap\&SS, 309, 167
\bibitem{FWZ04} Fan, Y.~Z., Wei, D.~M., \& Zhang, B.\ 2004, MNRAS, 354, 1031
\bibitem{JE91} Jones, F.~C. \& Ellison, D.~C. 1991, Space Sci.~Rev., 58, 259 
\bibitem{KSS00} Kaiser, C.~R., Sunyaev, R., \& Spruit, H.~C.\ 2000, A\&A, 356, 975 
\bibitem{KGGA00} Kirk, J.~G., Guthmann, A.~W., Gallant, Y.~A., \& Achterberg, A.\ 2000, ApJ, 542, 235
\bibitem{MP93} Melrose, D.~B., \& Pope, M.~H.\ 1993, PASAu, 10, 222 
\bibitem{MR99} M{\'e}sz{\'a}ros, P., \& Rees, M.~J.\ 1999, MNRAS, 306, L39 
\bibitem{PM94} Pope, M.~H., \& Melrose, D.~B.\ 1994, PASAu, 11, 175 
\bibitem{Rees78} Rees, M.~J.\ 1978, MNRAS, 184, 61P 
\bibitem{VV05} Virtanen, J.~J.~P, \& Vainio, R. 2005, ApJ, 621, 313

\end{thebibliography}

\end{document}